\documentclass[fleqn,aps,prl,nofootinbib,twocolumn,letterpaper,superscriptaddress]{revtex4-1}
\usepackage{dcolumn}
\usepackage{color}
\usepackage{graphicx}
\usepackage[colorlinks=true]{hyperref}
\usepackage{multirow}
\usepackage{float}
\usepackage{microtype}
\usepackage{booktabs}
\usepackage{amsmath,amsthm,amsfonts,amssymb,bm}
\usepackage{verbatim}
\usepackage{siunitx}
\usepackage{subcaption}
\usepackage{ragged2e}
\usepackage{ulem}
\usepackage{supertabular}
\usepackage{hyperref}
\usepackage[mathlines]{lineno}

\begin{document}

\title{Search for Solar Boosted Dark Matter Particles at the PandaX-4T Experiment}

\def\shKeyLab{School of Physics and Astronomy, Shanghai Jiao Tong University, Key Laboratory for Particle Astrophysics and Cosmology (MoE), Shanghai Key Laboratory for Particle Physics and Cosmology, Shanghai 200240, China}
\def\scKeyLab{Jinping Deep Underground Frontier Science and Dark Matter Key Laboratory of Sichuan Province}
\def\BUAA{School of Physics, Beihang University, Beijing 102206, China}
\def\BUAACenter{Peng Huanwu Collaborative Center for Research and Education, Beihang University, Beijing 100191, China}
\def\BUAALab{Beijing Key Laboratory of Advanced Nuclear Materials and Physics, Beihang University, Beijing, 102206, China}
\def\SCNT{Southern Center for Nuclear-Science Theory (SCNT), Institute of Modern Physics, Chinese Academy of Sciences, Huizhou 516000, China}
\def\USTClab{State Key Laboratory of Particle Detection and Electronics, University of Science and Technology of China, Hefei 230026, China}
\def\USTCdep{Department of Modern Physics, University of Science and Technology of China, Hefei 230026, China}
\def\BUAALab{International Research Center for Nuclei and Particles in the Cosmos \& Beijing Key Laboratory of Advanced Nuclear Materials and Physics, Beihang University, Beijing 100191, China}
\def\pku{School of Physics, Peking University, Beijing 100871, China}
\def\YaLongSD{Yalong River Hydropower Development Company, Ltd., 288 Shuanglin Road, Chengdu 610051, China}
\def\IAP{Shanghai Institute of Applied Physics, Chinese Academy of Sciences, 201800 Shanghai, China}
\def\CHEPpku{Center for High Energy Physics, Peking University, Beijing 100871, China}
\def\SDUdep{Research Center for Particle Science and Technology, Institute of Frontier and Interdisciplinary Science, Shandong University, Qingdao 266237, Shandong, China}
\def\SDUlab{Key Laboratory of Particle Physics and Particle Irradiation of Ministry of Education, Shandong University, Qingdao 266237, Shandong, China}
\def\UMD{Department of Physics, University of Maryland, College Park, Maryland 20742, USA}
\def\TDLee{New Cornerstone Science Laboratory, Tsung-Dao Lee Institute, Shanghai Jiao Tong University, Shanghai 201210, China}
\def\MESJTU{School of Mechanical Engineering, Shanghai Jiao Tong University, Shanghai 200240, China}
\def\SYU{School of Physics, Sun Yat-Sen University, Guangzhou 510275, China}
\def\SYUSFI{Sino-French Institute of Nuclear Engineering and Technology, Sun Yat-Sen University, Zhuhai, 519082, China}
\def\NKU{School of Physics, Nankai University, Tianjin 300071, China}
\def\YTU{Department of Physics, Yantai University, Yantai 264005, China}
\def\FDU{Key Laboratory of Nuclear Physics and Ion-beam Application (MoE), Institute of Modern Physics, Fudan University, Shanghai 200433, China}
\def\USST{School of Medical Instrument and Food Engineering, University of Shanghai for Science and Technology, Shanghai 200093, China}
\def\SJTUSC{Shanghai Jiao Tong University Sichuan Research Institute, Chengdu 610213, China}
\def\SPEIT{SJTU Paris Elite Institute of Technology, Shanghai Jiao Tong University, Shanghai, 200240, China}
\def\NNU{School of Physics and Technology, Nanjing Normal University, Nanjing 210023, China}
\def\SYSUzhuhai{School of Physics and Astronomy, Sun Yat-Sen University, Zhuhai 519082, China}
\def\CDUT{College of Nuclear Technology and Automation Engineering, Chengdu University of Technology, Chengdu 610059, China}
\def\Tsinghua{Department of Physics, Tsinghua University, Beijing 100084, China}
\def\THlab{Key Laboratory of Particle and Radiation Imaging (MoE), Tsinghua University, Beijing 100084, China}

\affiliation{\TDLee}
\author{Guofang Shen}\affiliation{\BUAA}
\author{Zihao Bo}\affiliation{\shKeyLab}
\author{Wei Chen}\affiliation{\shKeyLab}
\author{Xun Chen}\affiliation{\TDLee}\affiliation{\shKeyLab}\affiliation{\SJTUSC}\affiliation{\scKeyLab}
\author{Yunhua Chen}\affiliation{\YaLongSD}\affiliation{\scKeyLab}
\author{Zhaokan Cheng}\affiliation{\SYUSFI}
\author{Xiangyi Cui}\affiliation{\TDLee}
\author{Yingjie Fan}\affiliation{\YTU}
\author{Deqing Fang}\affiliation{\FDU}
\author{Zhixing Gao}\affiliation{\shKeyLab}
% ...
\collaboration{PandaX Collaboration}
\author{Haipeng An}\email[Corresponding author: ]{anhp@mail.tsinghua.edu.cn}\affiliation{\Tsinghua}\affiliation{\THlab}

\author{Haoming Nie}\affiliation{\Tsinghua}

\date{\today}

\begin{abstract}
We present a novel constraint on light dark matter utilizing $1.54$ tonne$\cdot$year of data acquired from the PandaX-4T dual-phase xenon time projection chamber. This constraint is derived through detecting electronic recoil signals resulting from the interaction with solar-enhanced dark matter flux. Low-mass dark matter particles, lighter than a few MeV/$c^2$, can scatter with the thermal electrons in the Sun. Consequently, with higher kinetic energy, the boosted dark matter component becomes detectable via contact scattering with xenon electrons, resulting in a few keV energy deposition that exceeds the threshold of PandaX-4T. We calculate the expected recoil energy in PandaX-4T considering the Sun's acceleration with heavy mediators and the detection capabilities of the xenon detector. The first experimental search results using the xenon detector yield the most stringent upper limits cross-section of $3.51 \times 10^{-39}~\mathrm{cm}^2$ at $0.08~\mathrm{MeV}$/$c^2$ for a solar boosted dark matter mass ranging from $0.02$ to $10~ \mathrm{MeV}$/$c^2$, achieving a 23 fold improvement compared with earlier experimental studies. 
\end{abstract}

\maketitle


 Although astrophysics and cosmology studies have accumulated ample evidence that much of the universe's mass is dark matter~(DM), its nature remains largely obscure~\cite{evidence2005,evidence2018,evidencePlanck2020}. 
The freeze-out mechanism proposes that dark matter was formed when particles frequently collided within a dense, high-temperature "soup" in the early universe. In this model, most particles initially exist as dark matter but later are destroyed. The remaining DM particles are usually classified as weakly interacting massive particles~(WIMPs)~\cite{Freezeout}. 
In recent decades, WIMPs have emerged as the predominant candidates for elucidating the phenomenon of missing mass, notwithstanding the lack of compelling evidence from direct detection experiments utilizing xenon or other materials~\cite{Direct-Detection2015,status2017, WIMPs2018, WIMPdirectdetection2019}. In recent years, xenon-based detectors such as PandaX-4T, LZ, and XENONnT have provided the most stringent limits on WIMPs–nucleon spin-independent interactions, pushing the cross-section down to $1.6 \times 10^{-47}~\mathrm{cm}^2$ for a DM mass of $40~\mathrm{GeV}/c^2$, $2.1 \times 10^{-48}~\mathrm{cm} ^2$ for $36~\mathrm{GeV}/c^2$, and $2.58 \times 10^{-47}~\mathrm{cm}^2$ for $28~\mathrm{GeV}/c^2$, ~respectively~\cite{Pandax2024Wimpresults, LZ2024Result, XenonNTFirstDarkMatter2023a}.
 These results present significant challenges to the mainstream supersymmetry framework~\cite{SUSY_supersymmetricdarkmatterlhc} in the WIMP sector. However, there remains considerable potential for further exploration in lower-mass regions.
 
The traditional method for directly detecting DM is to measure the elastic scattering between target nuclei and DM particles. Since cold DM in the halo is non-relativistic, the direct detection strategy is only sensitive to DM with masses exceeding sub-GeV due to the detector thresholds. In recent years, pushing the search to lower mass scales of sub-GeV or even sub-MeV DM, which are still untapped for promising thermal relic DM possibilities, has spurred many studies. Various approaches have been proposed, such as accessing the electron recoil (ER) channel~\cite{Essig2011, Emken2019, Essig2012, Essig2015, DarkSide2018, SuperCDMS2020, SENSEI2020, DAMIC2019,e-WIMP2021}, lowering the nuclear recoil (NR) detection threshold~\cite{Pirro2017, CRESST2019, CRESST2020, S2only2019, DarksideS2}, and utilizing the so-called Migdal effect~\cite{Ibe2017, Baxter2019, Essig2019, LUX2018, EDELWEISS2019, CDEX2019, Migdal2019}. 
Currently, considerable attention has been paid to specific processes that would boost the kinetic energy of DM particles~\cite{boosted2017,boosted_Sun2,boosted_cosmicray_cui,boosted_Self,boosted_inelastic,boosted_SuperK,boosted_Xe1t,boosted_Gra, Bringmann2020, boosted_blackholes, diurnalPRL}. For example, the search for DM boosted by cosmic rays presents constraints on DM-electron cross-sections ($\sigma_{e}$) for the unexplored DM mass from $10~\mathrm{eV}/c^2$  to $3~\mathrm{keV}/c^2$ in PandaX-4T~\cite{ShangXiaoFeng2024CR}. 

Another novel process called solar boosted dark matter~(SBDM) scenario has been proposed~\cite{boosted_Sun3,boosted_Sun, AnhaipengPRD} to open the window for sub-MeV DM detection. The Sun can act as an accelerator of DM particles via evaporation or scatterings with the thermal plasma in the core of the Sun where the temperature could reach keV~\cite{boosted_Sun_evaporation, boosted_Sun, boosted_Sun3, AnhaipengPRD}. 
In this letter, we focus exclusively on the generation and detection of SBDM through electron scattering~\cite{boosted_Sun3, AnhaipengPRD}, as the contribution of ions to MeV scale SBDM is negligible~\cite{SolarReflection_Emken2022}. 

The ambient DM with a typical velocity between $ 1 \times 10^{-3}$ and $ 2\times 10^{-3}~c$ would be attracted by the Sun via gravity when they pass nearby. The tremendous amount of energetic electrons in the Sun with a density $n_e$ at the level of $10 ^ {~25}~\text{electrons} /\mathrm{cm^{-3}}$ could effectively heat the DM particles and boost their kinetic energy to keV levels~\cite{AnhaipengPRD}. The SBDM could then escape the Sun and reach terrestrial detectors like PandaX-4T~\cite{Panda4T2021} leaving detectable signals that exceed the threshold with the same $\sigma_{e}$ amplitude on the bounded ~electrons of the xenon. Calculating the flux of SBDM emanating from the surface of the Sun is imperative. Subsequently, we will assess the event rate in the xenon detector based on the flux and $\sigma_{e}$ for different DM masses. 
The SBDM flux arriving at the earth can be estimated as ~\cite{boosted_Sun3}
\begin{equation}
\Phi_{\mathrm{boost}} \sim \frac{\Phi_{\text {halo }}}{4\pi} \times \begin{cases} S_g \frac{4 \pi}{3}\left(\frac{ R_{\text {core }}}{1 \mathrm{~A} .U.}\right)^2 \sigma_e n_e^{\text {core }} R_{\text {core }}, & \sigma_e \ll 1 \mathrm{pb} \\ S_g \pi \left(\frac{ R_{\text {scatt }}}{1 \mathrm{~A} . U .}\right)^2, & \sigma_e \gg 1 \mathrm{pb}\end{cases}
\label{equation1}
\end{equation}
where $\Phi_{\text{halo}}$ is the DM flux in the local Milky Way halo with a DM local density $\rho_{\mathrm{DM}} = 0.4~\mathrm{GeV}~\mathrm{cm}^{-3}$\cite{DMlocalDensity} and a typical DM velocity distribution, and $R_{\text{scatt}}$ is the scattering radius determined through the radius-temperature relation~\cite{Bahcall_2005}. $S_{\text{g}}$ indicates the gravitational focusing effect, which enhances the solar scatterings and has a value of $\sim$ O($10$) when $R_\text{scatt}\sim R_{\odot}$ ~\cite{boosted_Sun3}.  $~A.U.$ is the Sun-Earth distance. The $n_e^{\text {core }}$ is the electron number density within the radius of the Sun's core ($R_{\text {core }}$). The mean free path for a DM particle inside the core of the Sun is $(\sigma_e n_e^{\text {core }})^{-1}$ which would dramatically affect the calculations of the flux when $\sigma_{e} \ll 1~\mathrm{pb}$.

The possibility of multiple scattering should be considered in Eq.~\ref{equation1} but is difficult to treat analytically. So, a Monte Carlo simulation program~\cite{boosted_Sun3} is built to mimic the propagation of DM particles inside the Sun with an elaborate solar model taken from \cite{SSM_2009}. 
In the simulation, the perpendicular distance of halo DM particles to the Sun, which is known as impact parameter~$\rho$, is confined within $4R_{\odot}$ to incorporate gravitational focusing in the calculation of the SBDM. Beyond this range, only a small proportion of the slowest DM could be trapped by the gravity field, minimally contributing to the boosted flux. $F_{A_\rho}(E)$ is the simulated kinetic energy distribution of an SBDM particle, where $\int F_{A_\rho} (E) d E =1$, the DM mass, and $\sigma_{e}$ are two key input parameters for the simulation. The boosted flux observed at the detector’s location is illustrated as 
\begin{equation}
\frac{\mathrm{d} \Phi_{\text {boost}}}{\mathrm{d} E}=\Phi_{\text {halo }} \times \frac{A_\rho }{4 \pi(1 {~A} . {U}.)^2} \times F_{A_\rho}(E)
\end{equation}
where $A_\rho=\pi {\rho_\mathrm{{max}}^{2}}$ is the area where the DM particles could be boosted in the simulation and $\rho_{\mathrm{max}} = 4 R_{\odot}$.

\begin{figure}[!htbp]
\centering
\includegraphics[width=\linewidth]{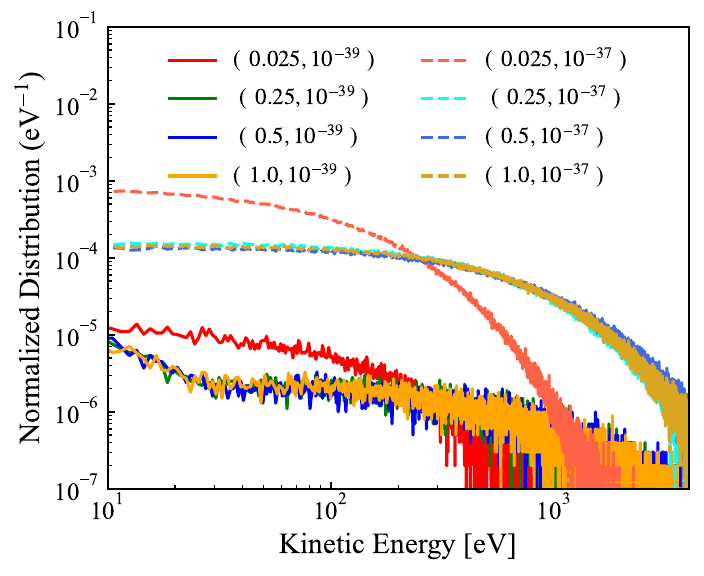}
\captionsetup{justification=RaggedRight, singlelinecheck=false}
\caption{SBDM kinetic energy spectrum. The first value in parentheses in the legend above represents the dark matter mass $m_{\mathrm{DM}}$ in units of MeV, and the value adjacent is $\sigma_{e}$ in units of $\mathrm{cm}^2$.}
\label{fig:1}
\end{figure}

Taking into account $m_\mathrm{DM} = 0.025,~0.25,~0.5$, and $1.0~\mathrm{MeV}$, the SBDM spectra, with $\mathrm{\sigma_{e}}$ = $10^{-39}$ and $10^{-37}~\mathrm{cm^{2}}$, are shown in Fig.~\ref{fig:1}. The kinetic energy transfer efficiency reaches a maximum when the masses of the elastically scattering particles are equal. The spectra for 
$m_\mathrm{DM}$ = 0.5~MeV are harder than those for the other three DM masses. 

According to the DM-electron scattering analysis~\cite{Essig2012}, the velocity-averaged electron recoil differential cross-section for ionization from an initial atomic state
with principal and angular quantum numbers $n$ and $l$ can be expressed as
\begin{equation}
\frac{d\left\langle\sigma_{n l}\right\rangle}{d E_{R}}=\frac{1}{E_{R}}\frac{\sigma_e}{8 \mu_{\mathrm{DM}, \mathrm{e}}^2} \int d q q\left|f_{n l}\left(q, p_e^{\prime}\right)\right|^2\left|F_{\mathrm{DM}}(q)\right|^2 \eta\left(v_{\min}\right)
\end{equation}
where $\mu_{\mathrm{DM},\mathrm{e}}$ is the DM-electron reduced mass, and $E_{R}$ is the recoil energy. The DM form factor $F_{\mathrm{DM}}(q)$ can take different values depending on the momentum transfer $q$. The atomic form factor,
$|f_{n l}(q,p_e^{\prime})|^2$, describes the strength of the ionization process from the atomic state ($n,~l$). In this work, we consider $F_{\mathrm{DM}}(q) = 1$~\cite{AnhaipengPRD}, which means that the interaction between SBDM particles and electrons is contact. $\eta(v_{\min })$ is the flux-average of squared inverse speed for the SBDM with a velocity larger than $v_\mathrm{min}$ that could generate a minimum recoil. 

There is a framework~\cite{Essig2015} built to calculate sub-GeV dark matter scattering off electrons with
semiconductor targets.
Our entire calculation is based on the framework to calculate the DM-electron scattering rate, but with the following modifications.
First, we use xenon as the target; second, we sum the electron recoiling and deexcitation photon energy from different electron shells in xenon; last,
the atomic form factor is taken from Ref.~\cite{FormFactorEmken}, also used in cosmic-ray boosted DM research in PandaX~\cite{ShangXiaoFeng2024CR}.

Next, we calculate the differential recoil rate $R$   
\begin{equation}
\frac{d R}{d E_R}=n_t \times \sum_{n l}  \int dE \frac{d \Phi_{\text {boost}}}{d E} \frac{d\left\langle\sigma_{n l}\right\rangle}{d E_{R}}
\end{equation}
where $n_{t}$ is the electron number density in the xenon target. ${d\langle\sigma_{n l}\rangle}/{d E_{R}}$ is the differential scattering cross-section, and the ${d \Phi_{\text{boost}}}/d E$ is the SBDM flux at the detector’s location~\cite{boosted_Sun3}.

PandaX-4T~\cite{Panda4T2021} is located in the China Jinping Underground Laboratory (CJPL)~\cite{CJPL1, CJPL2}. The central detector is a dual-phase time projection chamber~(TPC) with $3.7$ tonnes of liquid xenon in the sensitive region. The detector is defined by $24$ reflective polytetrafluoroethylene~(PTFE) panels, separating $1185$ mm between opposing panels. The internal electric fields are established by a cathode grid at the bottom, along with a gate and anode mesh positioned just below and above the liquid level, maintaining vertical spacings of $1185$ and $10$ mm, respectively. In addition, two arrays of $169$ and $199$ three-inch Hamamatsu R11410-23 photomultiplier tubes (PMT) $169$ and $199$ are installed on the top and bottom of the TPC.
 
 Electron recoil energy is converted into prompt scintillation ($S1$) in liquid xenon and delayed proportional electroluminescence photons ($S2$) in gaseous xenon. The signals $S1$ and $S2$ are collected by the PMT arrays and used to reconstruct the energy and three-dimensional position. To enhance our understanding of the detector response, we utilize two primary calibration methods: the calibration of electronic and nuclear recoil events. The ER response is calibrated by the $^{220}$Rn~\cite{Ma_2020} and $^{83m}$Kr~\cite{ZhangDan_2022} and other radioactive sources such as $^{60}$Co and $^{137}$Cs. The Am-Be neutron source and the Deuterium-Deuterium neutron generator are used for NR calibration. More details are given in~\cite{Pandax2024Wimpresults}.

The electron equivalent energy $E_{\rm{re}}$ of each event in the detector is reconstructed as
\begin{equation}
E_{\rm{re}}= W \times \left(\frac{S1}{{\mathrm{PDE}}} +\frac{S2_{b}}{{\mathrm{EEE}}\times {\mathrm{SEG}_{b}}}\right)  
\end{equation}
 where $W$ is the average work function of liquid xenon, with a value of 13.7~eV from ~\cite{Workfunction}. The $S2_{b}$ is the bottom-only $S2$. The use of $S2_{b}$ is to avoid the effects of
saturation and dead channels of the top PMTs.
 PDE, EEE, and $\mathrm{SEG}_b$ represent the photon detection efficiency, electron extraction efficiency, and single electron gain, respectively. The $\mathrm{SEG}_{b}$ here already involve the PDE for S2 and are obtained by measuring the smallest recorded $S2_{b}$ signals~\cite{Panda4T2021}. The values of PDE and EEE are predetermined by calibrating the monoenergetic electronic recoil peaks: 41.5 keV ($^{83m}$ Kr), 164 keV ($^{131m}$Xe), 236 keV ($^{129m}$Xe), and 408 keV ($^{127}$Xe).  

\begin{figure}[!htbp]
\centering
\includegraphics[width=\linewidth]{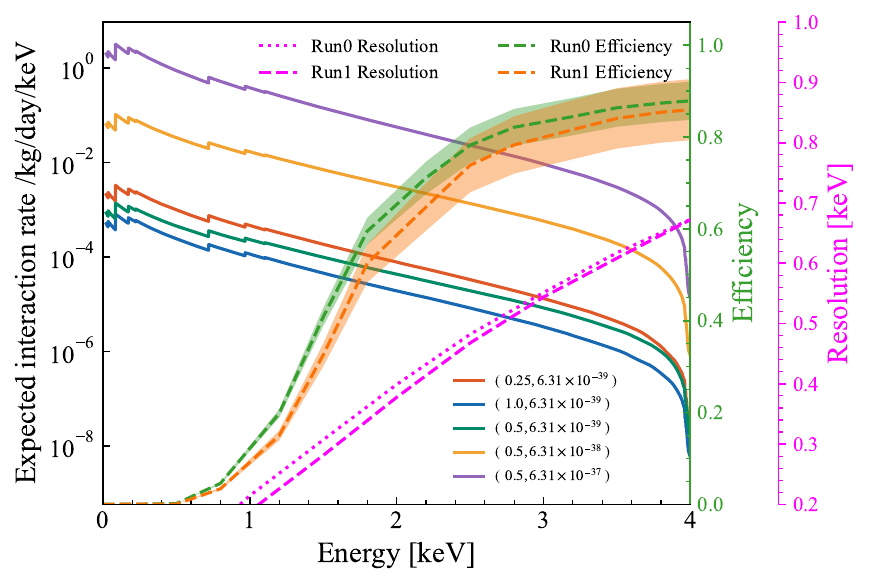}
\captionsetup{justification=RaggedRight, singlelinecheck=false}
\caption{The expected interaction rate of SBDM in the xenon detector with $m_\mathrm{DM}$ = 0.25~MeV, 0.5~MeV, and 1.0~MeV for $\sigma_{e} = 6.31 \times 10^{-39}~\mathrm{cm}^2$. For $m_\mathrm{DM}$ = 0.5 MeV, the expected interaction rate with $\sigma_{e}$ from $6.31 \times 10^{-39}~\mathrm{cm}^2$ to $6.31 \times 10^{-37}~\mathrm{cm}^2$ cross-sections is also shown. The dashed green and dashed orange lines represent the total efficiencies of Run0 and Run1, and the shield bands illustrate the efficiency uncertainties. The dotted and dashed magenta curves are the energy resolution of Run0 and Run1, respectively. With interaction rate, efficiency and energy resolution convolved, the corresponding reconstructed signal spectrum could be derived and used in the following analysis.
}
\label{fig:2}
\end{figure}

Fig.~\ref{fig:2} illustrates the expected interaction rate for different DM masses with representative cross-sections and the total efficiencies in PandaX-4T. The total efficiencies are estimated using both the data-driven and waveform-simulation method~\cite{LiJiaFu_2024, P4SignalModel}. Furthermore, the energy resolution curves of Run0 and Run1 are given in Fig.~\ref{fig:2}. The structure of the energy spectrum is attributed to interactions between the SBDM and different shell electrons of the xenon. The recoil energy could reach four keV, beyond the PandaX-4T experiment threshold~\cite{Panda4T2021, Pandax2024Wimpresults}. 

This analysis uses the same data sets from Run0 and Run1 as Ref.~\cite{Pandax2024Wimpresults}, including 11 subsets segmented according to slightly different electric field configurations and background levels. The fiducial masses are $2.38\pm 0.01$ tonne for Run0 and $2.47\pm 0.02$ tonne for Run1. The total exposure is $1.54$ tonne$\cdot$year.

In the search for SBDM, the signals are expected to manifest themselves as electronic recoil events, in contrast to conventional WIMP signals. Consequently, an accurate estimation and characterization of all ER backgrounds is essential for the statistical analysis and interpretation of SBDM signals. 
The energy of SBDM signals ranges from $0$ to $4$ keV. To enhance the measurement of the ER background from ${}^{222}$Rn, we choose the final region-of-interest~(ROI) from 0 to 30~keV. Furthermore, a threshold of 99.5\% ER acceptance cut was implemented to exclude most NR events arising from solar $^8$B neutrinos and neutron interactions with the xenon nucleus as well as radon plateout on the surface of the PTFE reflector of the PandaX-4T TPC. This cut is derived from the simulated data of the PandaX-4T signal model~\cite{P4SignalModel}. Finally, we consider the primary ER backgrounds, including tritium, $^{85}$Kr, $^{214}$Pb, $^{212}$Pb, $^{127}$Xe, $^{136}$Xe, $^{124}$Xe, neutrino, material, and accidental events. A comprehensive examination of each component is presented in~\cite{axionzeng2024}.  

\begin{table*}
    \centering
    \resizebox{\linewidth}{!}{
    \setlength{\tabcolsep}{7mm}{
    \begin{tabular}{l|ll|ll}
    \toprule[1.5pt]
     
 Run & \multicolumn{2}{c}{Run0} \vline & \multicolumn{2}{c}{Run1}\\
    \midrule[1pt]
 Exposure & \multicolumn{2}{c}{198.9~tonne$\cdot$day} \vline & \multicolumn{2}{c}{363.3~tonne$\cdot$day} \\
      \midrule[1pt]
       & Expected & Fitted & Expected & Fitted \\
    \midrule[1pt]
 Tritium  & --- & 575.1 $\pm$ 32.8 & --- & $115.2 \pm 32.0$ \\
    $^{214}$Pb & 327.2 $\pm$ 18.8 & 328.1 $\pm$ 17.1& $724.2 \pm 61.5$& $702.6 \pm 28.4$\\
    $^{212}$Pb & 57.8 $\pm$ 14.7& $57.3 \pm 14.1$ & $103.3 \pm 26.9$& $96.6 \pm 23.8$ \\
    $^{85}$Kr & 94.2 $\pm$ 47.3 & $87.5 \pm 31.2$& $308.1 \pm 95.2$ & $272.7 \pm 59.0$\\
    Material & 49.4 $\pm$ 3.3 & $49.5 \pm 3.1$ & $111.7 \pm 9.9$ & $106.0 \pm 7.8$\\
    $^{136}$Xe  & 36.9 $\pm$ 2.5 & $36.9 \pm 2.4$ & $66.2 \pm 5.9$ & $62.4 \pm 4.6$ \\
    $^{127}$Xe & 6.1 $\pm$ 0.3 & $7.6 \pm 0.4$ & $0.0 \pm 0.0$ & $0.0 \pm 0.0$\\
    $^{124}$Xe & 2.3 $\pm$ 0.6 & $2.3 \pm 0.4$ & $4.0 \pm 1.1$ & $3.9 \pm 1.1$\\
 Solar $\nu$ & 43.0 $\pm$ 4.6 & $42.9 \pm 4.5$& $76.8 \pm 9.4$& $72.7 \pm 8.1$\\
 Accidental & 7.6 $\pm$ 2.4 & $7.7 \pm 2.2$ & $7.1 \pm 2.3$ & $6.8 \pm 2.0$\\
  Signal &  --- &{$1.7_{-1.7}^{+3.1}$} & --- &{$2.5_{-2.5}^{+4.6}$}\\
  \midrule[1pt]
 Total fitted &  \multicolumn{2}{c}{1196.5$\pm$ 32.7}\vline & \multicolumn{2}{c}{1439.4 $\pm$ 36.2}\\
 \midrule[1pt]
  Observed & \multicolumn{2}{c}{1197} \vline & \multicolumn{2}{c}{1431} \\
 \bottomrule[1.5pt]
 \end{tabular}
 }
 }
 \captionsetup{justification=RaggedRight, singlelinecheck=false}
\caption{Summary of the expected background and signal, and the best fit of the background plus signal hypothesis for each component in Run0 and Run1. Systematic uncertainties associated with the signal model and detection efficiencies are incorporated into the reported uncertainties.}
\label{tab:bkg_signal_fit}
\end{table*}

\begin{figure}[!htbp]
\centering
\includegraphics[width=\linewidth]{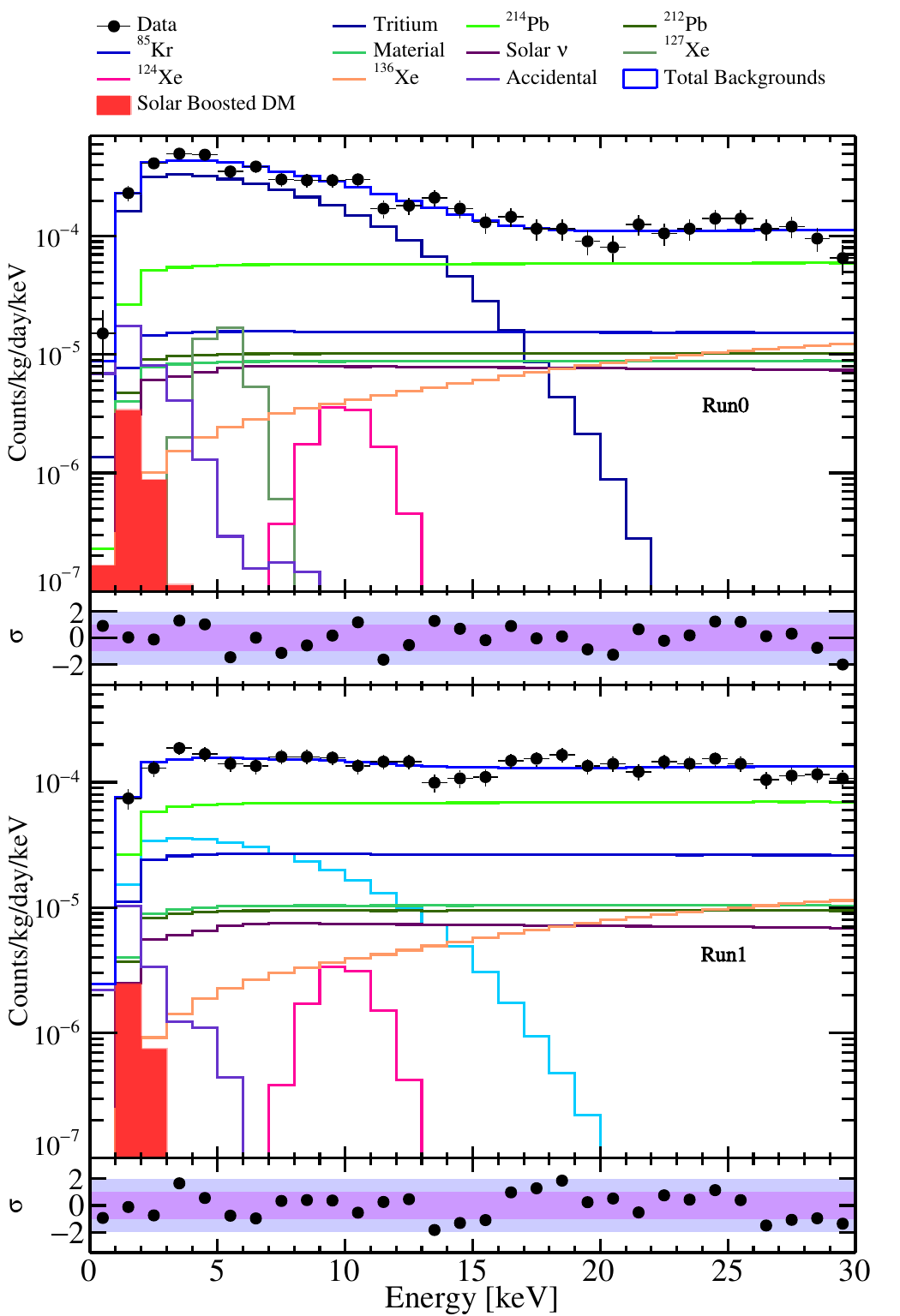}
\captionsetup{justification=RaggedRight, singlelinecheck=false}
\caption{Best combined fit results of Run0 and Run1 data with background-plus-signal hypothesis. Error bars are statistical uncertainties.}
\label{fig:3}
\end{figure}

Data from the PandaX-4T experiments, specifically the
1197 (Run0) and 1431 (Run1) electron recoil (ER) events, have been meticulously selected for this analysis. A profile likelihood ratio~(PLR) approach~\cite{DMStatWhitePaper} is used in this analysis with a one-dimensional signal energy spectrum based on HistFitter~\cite{HistFitterBaak2015}. In this analysis, both background-only and background-plus-signal hypotheses are tested. Background contributions are well aligned with the expected values, and all nuisance parameters remain within $\pm 1\sigma$ of the input values. The best fit of the SBDM result is illustrated in Fig.~\ref{fig:3} and the expected events from each component are summarized in Table~\ref{tab:bkg_signal_fit}. Ultimately, no significant SBDM signal is detected above the background.

\begin{figure}[!htbp]
    \centering
    \includegraphics[width=\linewidth]{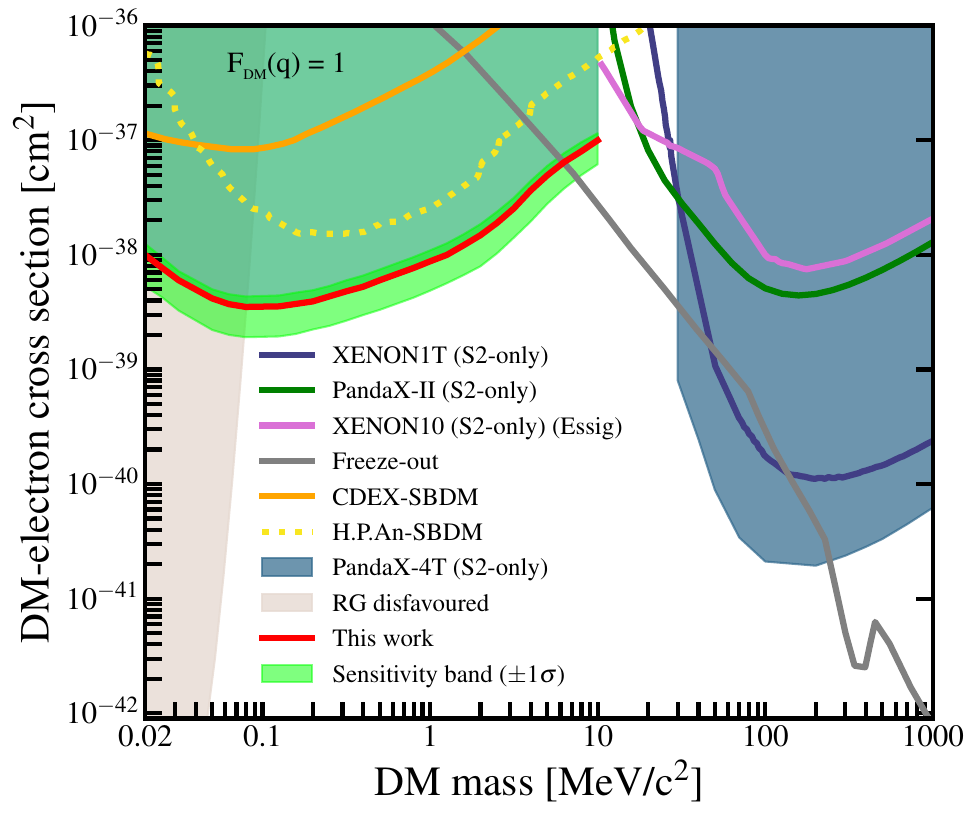}
    \captionsetup{justification=RaggedRight, singlelinecheck=false}
    \caption{ 90\% confidence level~(C.L.) exclusion upper limits~(red line) and exclusion region~(lightgreen region)~for the DM-electron cross-section are presented for SBDM. The light gray band is stellar cooling constraints from red giant (RG) stars~\cite{RG}. The gray curve denotes the freeze-out DM mechanism~\cite{Planck2015_freezeout}, while the dashed yellow line indicates the SBDM limits from Ref.~\cite{boosted_Sun3}. Additionally, the orange line is from the CDEX experimental result~\cite{CDEXSRDM}, and the direct constraints based on S2 analysis from Ref.~\cite{PandaX-4TS2only, PandaX-IIcc, Xenon1TS2only, Essig2012} are also overlaid at the right side.}
    \label{fig:4}
\end{figure}

The 90\% confidence level~(C.L.) two-side upper limit exclusion~(solid red line) of the SBDM parameter space, along with the $\pm 1\sigma$ sensitivity~(green band), is presented in Fig.~\ref{fig:4}.
For comparison, recent results from the phenomenological study~\cite{boosted_Sun3}, the SBDM analysis utilizing the HPGe detector from CDEX~\cite{CDEXSRDM}, and the constraints derived from cosmological and astrophysical observables related to freeze-out DM~\cite{Planck2015_freezeout}, and other analyses of sub-GeV DM~\cite{PandaX-4TS2only, PandaX-IIcc, Xenon1TS2only, Essig2012} are overlaid. Our results cover a large region from 20~keV/$c^2$ 
to 10~MeV/$c^2$, with cross-sections ranging between $10^{-39}~\mathrm{cm^{2}}$ and
$10^{-38}~\mathrm{{cm}^2}$. The lowest 90\% confidence level on the SBDM cross-section is $3.51\times 10^{-39}~\mathrm{cm}^2$ for an SBDM mass of $0.08~\mathrm{MeV}$$/c^2$. Our results provide the most stringent constraints, achieving a 23-fold improvement compared to recent experimental studies conducted by CDEX~\cite{CDEXSRDM}.

In summary, we have performed the direct detection of DM-electron scattering by searching for an energetic
SBDM flux produced through rescattering in the
Sun with the 1.54~tonne$\cdot$year data of Run0 and Run1 from PandaX-4T. This leads to a new sensitivity at the sub-pb level for light DM in the sub-MeV mass range. No significant dark matter signals are identified above the expected background. A new upper limit exclusion is established for the SBDM-electron interactions with heavy mediators, robustly excluding sub-MeV DM with a scattering cross-section with electrons within $10^{-39}~\mathrm{{cm}^{2}}$ to $10^{-38}~\mathrm{{cm}^{2}}$. The well-known freeze-out mechanism below the $10~\mathrm{MeV}$$/c^2$ region could be excluded. Combined with analyses such as ionization signals~\cite{PandaX-4TS2only}, 
The future upgraded PandaX-4T detectors, with significantly enhanced sensitivity and optimized S2-only event detection thresholds~\cite{PandaX4t2PandaX-xT}, will explore the narrow unconstrained parameter window in the mass range of $10-30~\mathrm{MeV}$$/c^2$.


 
We thank Josef Pradler and Shao-Feng Ge for their helpful discussions. This project is supported in part by grants from National Science Foundation of China (Nos. 12090060, 12090061, 12105008, 12435007, U23B2070), a grant from the Ministry of Science and Technology of China (Nos. 2023YFA1606200, 2023YFA1606201, and by Office of Science and Technology, Shanghai Municipal Government (grant No. 22JC1410100, 21TQ1400218), the National Natural Science Foundation of China under Grant No. 12435007. We thank for the support by the Fundamental Research Funds for the Central Universities. We also thank the sponsorship from the Chinese Academy of Sciences Center for Excellence in Particle Physics (CCEPP), Hongwen Foundation in Hong Kong, New Cornerstone Science Foundation, Tencent Foundation in China, and Yangyang Development Fund. Finally, we thank the CJPL administration and the Yalong River Hydropower Development Company Ltd. for indispensable logistical support and other help. 

\normalem
\bibliography{apssamp}
\clearpage
\end{document}